\documentclass[twocolumn,showpacs,preprintnumbers,amsmath,amssymb]{revtex4}

\usepackage{graphicx}
\usepackage{dcolumn}
\usepackage{bm}
\usepackage{colordvi}
\usepackage{epsfig}
\usepackage{graphicx}
\usepackage{color}

\newcommand{\kap}{\boldsymbol{\kappa}}
\newcommand{\rb}{{\bf r}}
\newcommand{\bp}{\boldsymbol{\partial}}
\newcommand{\qb}{{\bf q}}
\newcommand{\kb}{{\bf k}}
\newcommand{\pb}{{\bf p}}

\begin{document}

\title{Dense Colloidal Suspensions under Time-Dependent Shear}

\author{J.~M.~Brader$^1$, T.~Voigtmann$^2$, M.~E.~Cates$^2$ and M.~Fuchs$^1$}
\affiliation{
$^1$Fachbereich Physik, Universit\"at Konstanz, D-78457 Konstanz, Germany
\\
$^2$SUPA, School of Physics, The University of Edinburgh, 
Mayfield Road, Edinburgh EH9 3JZ, UK
}


\begin{abstract}
We consider the nonlinear rheology of dense colloidal suspensions under a time-dependent simple 
shear flow. 
Starting from the Smoluchowski equation for interacting Brownian particles 
advected by shearing (ignoring fluctuations in fluid velocity) we develop a formalism which 
enables the calculation of time-dependent, far-from-equilibrium averages. 
Taking shear-stress as an example we derive exactly a generalized Green-Kubo relation, 
and an equation of motion for the transient density correlator, involving a 
three-time memory function. 
Mode coupling approximations give a closed constitutive equation yielding the time-dependent 
stress for arbitrary shear rate history.
We solve this equation numerically for the special case of a hard sphere glass subject 
to step-strain.          
\end{abstract}

\maketitle

The constitutive equations of a material relate its stress tensor 
$\boldsymbol{\sigma}(t)$ 
at time $t$ to its flow history, described by a strain-rate tensor $\kap(t'\!<\!t)$. 
Dense suspensions of colloidal particles, 
close to the glass transition, show strongly viscoelastic behaviour and nonlinear shear response; 
their constitutive equations must be highly nontrivial \protect\cite{expts}. In this paper, we report 
first-principles work aimed at elucidating these equations theoretically.
\par
Three alternative approaches to the rheology of glassy materials have recently been considered.
The phenomenological soft glassy rheology (SGR) model \protect\cite{sgr}  mimicks the glass transition 
by a one-particle hopping dynamics in a well chosen distribution of trap-depths \protect\cite{bouchaud}. 
While the predictions of SGR are broadly consistent with experiments on many non-ergodic 
soft materials \protect\cite{cowpaste} it does not capture the 
discontinuous jump in yield stress on glass formation observed in experiments on  
colloidal suspensions of hard spheres \protect\cite{expts}. 
The same is true of  spin glass approaches which describe a different phenomenology akin to 
`power-law yielding 
materials' \protect\cite{spingl}. 
The observed yield-stress is captured however by a first-principles approach 
to colloid rheology, based on Mode Coupling Theory (MCT), 
which has recently been formulated for systems under steady shear
\protect\cite{fuchs_cates_PRL}. 
The MCT has had considerable semiquantitative success in accounting for the 
interaction dependence of the static glass transition 
\protect\cite{goetze_sjoegren,pham2002} and the time dependence of light scattering 
correlators \protect\cite{vanmegen} from microscopic starting points. 
MCT also gives sensible predictions for the viscoelastic spectrum $G^*(\omega)$ as measured in 
linear rheology \protect\cite{MCTequations,gstar}.

In this Letter we develop a first-principles description of the 
far-from-equilibrium states of dense colloids under prescribed time dependent shear flow. 
For a system of interacting particles advected by the imposed strain rate $\kap$ (this neglects 
velocity fluctuations and hence hydrodynamic interactions \protect\cite{fuchs_cates_PRL}), 
we begin by developing a formalism which facilitates calculation of general time dependent 
averages.
Using this formalism we derive formally exact generalized Green-Kubo equations, 
taking the shear stress as an example.
These expressions can be approximated in terms of transient two-time density correlators. 
We find an exact equation of motion for the correlator which, 
for general strain rate, displays a remarkable three-time structure to the 
memory function. 
Finally, we make an MCT-based closure of this equation. 
To show that the theory yields sensible predictions for strongly time-dependent flows, we consider 
step-strain 
as a specific example. 
Our approach is valid for all homogeneous and incompressible flows which satisfy 
$\kap(t)\!\cdot\!\kap(t)\!\!=\!\!0$ 
(this enforces homogeneous states with translationally invariant spatial correlations). 
For clarity of presentation we focus on the case of simple shearing 
with fixed axes (velocity along $x$, gradient along $y$), and an arbitrary time-dependent strain 
rate $\dot\gamma(t)$, so that $\kap_{ij} = \dot\gamma(t)\delta_{ix}\delta_{jy}$.

Our findings highlight the formal importance of `integration through 
transients' (ITT) \protect\cite{fuchs_cates_PRL} in 
preparing the best ground for judicious application of  MCT. Indeed, a somewhat simpler 
MCT-inspired approach to colloid rheology was developed in \protect\cite{miyazaki_reichman,miyazaki_reichman2}, 
which for fluid states in steady shear gives broadly similar results to those of \protect\cite{fuchs_cates_PRL}. 
However, its recent ({\em ad-hoc}) extension to time dependent flows \protect\cite{miyazaki_reichman2} 
gives results quite different from ours; in particular, there is no sign of a third time in the 
memory function which we show to be an exact consequence of the Smoluchowski equation and which is
preserved in our approximations. 
This feature (also missed in another recent simplified MCT approach \protect\cite{kss}) 
turns out to be crucial in capturing the nonlinear response, even for simple time-dependent 
flows. 
Our findings should also help guide the development of `fully schematic' 
single-mode models. 
These models can address 
physics not considered here, such as shear thickening \protect\cite{holmes}, so their extension to 
time-dependent flows is highly desirable.
In future, such models may suggest improvements to the MCT factorizations used here, 
iterating towards a fully predictive theory of colloid rheology.

We start with a system of $N$ spherical Brownian particles of diameter $d$ 
interacting via internal 
forces ${\bf F}_i$, $i=1,\cdots,N$, and dispersed in a solvent 
with a specified time dependent velocity profile ${\bf v}(\rb,t)=\kap(t)\cdot\rb$.
The distribution function of particle positions evolves according to the 
Smoluchowski equation \protect\cite{dhont,doi_edwards}
\begin{eqnarray}
\partial_t \Psi(t) &=& \Omega(t) \Psi(t) \notag\\
\Omega(t) &=& \sum_i \bp_i\cdot[\bp_i - {\bf F}_i - \kap(t)\cdot\rb_i],
\label{smol}
\end{eqnarray}
where $\Omega(t)$ is the Smoluchowski operator and we have introduced dimensionless 
variables for length, energy and time, $d=k_BT=D_0=1$. 
Translational invariance of the sheared system leads to a coupling between a 
density fluctuation 
with pre-advected wavevector $\qb(t,t')=\qb - \qb\cdot\int_{t'}^tds\,\kap(s)$
at time $t'$ and another with wavevector 
$\qb$ at later time $t$. 
Wavevector advection is how  strain enters our 
formalism and accounts for the affine part of the particle motion in the imposed 
flow.
\par 
Integrating the Smoluchowski equation we obtain the following formal solution 
for the distribution function
\begin{equation}
\Psi(t) = \Psi_e + 
\int_{-\infty}^{t}\!\!\! dt'\; \Psi_e\;{\rm Tr}\{\kap(t)\hat{\boldsymbol{\sigma}}\}\; 
e_-^{\int_{t'}^t ds\,\Omega^{\dagger}(s)},
\label{distribution}
\end{equation}
where $\hat{\boldsymbol{\sigma}}$ is the zero wavevector limit of the potential 
part of the stress tensor and 
$\Psi_e$ is the equilibrium distribution function. 
The adjoint Smoluchowski operator can be obtained by partial integration
and is given by 
$\Omega^{\dagger}(t) = \sum_i [\bp_i + {\bf F}_i + \rb_i\cdot\kap^T(t)]\cdot\bp_i$.
The time ordered exponential $e_-$ imposes that later times appear on the right 
and arises because the operator $\Omega(t)$ does not commute with itself at different 
times.
Eq. (\ref{distribution}) generalizes the ITT method developed in \protect\cite{fuchs_cates_PRL} 
and is to be used with the understanding that functions to 
be averaged are placed to the right of the operators and then integrated over 
particle coordinates. 
A general function $f$ of the phase space coordinates thus has the time dependent 
average  
$\langle f \rangle(t) 
= \langle f \rangle + \int_{-\infty}^{t}dt' 
\langle{\rm Tr}\{\kap(t)\hat{\boldsymbol{\sigma}}\}\exp_-[\int_{t'}^t ds\Omega^{\dagger}(s)]
f \rangle$
with respect to the distribution function (\ref{distribution}), 
where $\langle\cdot\rangle$ indicates averaging with respect to the equilibrium 
distribution. 
By choosing $f=\sigma_{xy}/V$ \protect\cite{normalstress} 
we obtain an 
exact generalized Green-Kubo relation for the time dependent shear stress (in volume $V$)
\begin{equation}
\sigma(t) = \int_{-\infty}^{t}\!\!\! dt'\, \dot\gamma(t')\!\left[\frac{1}{V} \langle  
\sigma_{xy} e_-^{\int_{t'}^t ds\Omega^{\dagger}(s)} \sigma_{xy}\rangle\right],
\label{exactstress}
\end{equation}
where the factor $[\cdot]$  can be identified formally as $G(t,t',[\dot\gamma])$, a time-dependent 
shear modulus. Replacing $\Omega^{\dagger}(t)$ with the quiescent-state operator recovers linear 
response.
Eq.~(\ref{exactstress}) opens a route to calculate $\sigma(t)$ for a given flow history 
$\dot\gamma(t)$. 
The ITT method based on Eq.~(\ref{distribution}) also yields expressions for 
correlators, distorted structure factors and susceptibilities which will be detailed elsewhere.
\par 
To approximate our formally exact result (\ref{exactstress}) we now project
onto densities $\rho_{\qb}=\sum_{i}e^{i\qb\cdot\rb_i}$ and density pairs (given by the square of 
the density in real space) \protect\cite{goetze_sjoegren,fuchs_cates_PRL}.
This physical approximation amounts to assuming that these are the only slow variables, sufficient 
to describe the relaxation of the local structure in the glassy regime. 
The resulting shear stress is given by
\begin{eqnarray}
\sigma(t)\!=\!\!\int_{-\infty}^{t} \!\!\!\!\!\!\!\!dt'\dot{\gamma}(t')\!\! \int\!\!\! 
\frac{d{\bf k}}{16\pi^3}\frac{k_x^2 k_yk_y(t,t')}{k k(t,t')}\frac{S'_{k}S'_{k(t,t')}}{S^2_{k(t,t')}}
\Phi^2_{\bf k}(t,t'), 
\label{MCTstress}
\end{eqnarray}
where $\Phi_{\bf k}(t,t')$ is the transient density correlator 
which (in the absence of time translational invariance) is  a function of two times, and 
$S'_k = dS_k/dk$ with $S_k$ the equilibrium static structure factor. 
The projection onto density pairs means that the interparticle forces ${\bf F}_i$ are fully 
determined from $S_k$ and density fluctuations. 
This MCT-based approximation is well-tested, although if the equal-time structure under shear 
deviates strongly enough from $S_k$ to enter an anharmonic regime, improvements to it may 
be needed \protect\cite{fuchs_cates_PRL,holmes}.
In Eq.~(\ref{MCTstress}), the term $\Phi^2_{\bf k}(t,t')$ can be viewed as the 
`survival probability' to time $t$ of a stress contribution created by an initial step-strain 
applied at earlier time $t'$; the remaining factor is the stress per unit initial strain.
The transient density correlator required in Eq.~(\ref{MCTstress}) is defined as    
$
\Phi_{\bf q}(t,t') = \langle \rho^*_{\bf q}\,
\exp_-[\int_{t'}^{t}ds\,\Omega^{\dagger}(s)]\,\rho_{{\bf q}(t,t')} \rangle
/(NS_{q})
$
and is a key quantity within our approach.
Note that it only contains 
information on the strain accumulated between the two correlated 
times $t', t$ and is independent of the strain history for times earlier 
than $t'$.  
\par
Equation (\ref{MCTstress}) gives the stress in terms of 
$\Phi_{\bf k}(t,t')$ which is itself dependent on flow history. 
Using Zwanzig-Mori type projection operator manipulations and applying the theory of Volterra 
integral equations, we obtain the following formally exact results:
\begin{eqnarray}
\frac{\partial}{\partial t} \Phi_{\bf q}(t,t_0) &+& \Gamma_{\bf q}(t,t_0)\bigg( 
\Phi_{\qb}(t,t_0)   
\label{eom}
\\
&+& \int_{t_0}^t dt' m_{\qb}(t,t',t_0) \frac{\partial}{\partial t'} \Phi_{\qb}(t',t_0)
\bigg) =0, \notag
\\
\label{initialdecay}
\Gamma_{\bf q}
(t,t_0) 
&=& -\frac{1}{S_{\bf q}}
\langle \rho^*_{\bf q} \Omega_s(t,t_0)\rho^s_{\bf q} \rangle \label{exactdecay}
\\
&=& \frac{q^2}{S_q}
-\qb\cdot\kap_{tt_0}\cdot\qb \frac{1+S_q}{S_q}
+|\qb\cdot\kap_{tt_0}|^2
\notag
\\
m_{\bf q}
(t,t',t_0) 
&=& \!\!\frac{-
\langle
\rho^*_{\bf q} \Omega_s(t',t_0) U(t,t',t_0)
Q_s\Omega_s(t,t_0)\rho^s_{\bf q}
\rangle}
{
S_{q} \Gamma_{\bf q}(t',t_0) \Gamma_{\bf q}(t,t_0)
}. 
\notag\\
\label{exactmemory}
\end{eqnarray}
In the equation of motion (\ref{eom}) for $\Phi_{\bf k}(t,t')$, the initial decay rate 
$\Gamma_{\bf q}(t,t')$ describes Taylor dispersion 
which enhances diffusion in the direction of flow \protect\cite{dhont}. 
For our chosen flow geometry, this can be calculated explicitly as given in Eq.~(\ref{exactdecay}),
where $\kap_{tt_0}\!=\!\int_{t_0}^{t}ds\,\kap(s)$ is the shear strain accumulated between $t_0$ 
and $t$.
The memory function $m_{\qb}(t,t',t_0)$ in Eq.~\ref{eom} describes competition between 
shearing and the cage effect responsible for slow structural relaxation; strikingly, this is a 
function of three times, not two. 
It is useful to interpret $m_{\qb}(t,t',t_0)$ as describing the decay 
of memory between times $t'$ and $t$, in the presence of shear, allowing for the coupling to stress 
in the system that is still relaxing from the strain accumulated between $t_0$ and $t'$. 
The time $t_0$ enters the theory in a parametric fashion and is quite distinct in 
character from the two later times. 
Eq.~(\ref{exactmemory}) for $m_{\qb}$ involves the propagator
$U(t,t',t_0) = \exp_-\int_{t'}^t ds' Q_s\Omega^{\rm irr}_s(s',t_0)$, 
where $\Omega^{\rm irr}_s(t,t')$ is the single-particle-irreducible 
operator \protect\cite{fuchs_cates_PRL,ch} and $Q_s$ is an equilibrium projector
orthogonal to density fluctuations.
In deriving these formal results we introduced $\rho_{\qb}^s=e^{i\qb\cdot\rb_s}$, 
the density of a single tagged particle, whose motion is described 
by  
$\Omega_s(t,t_0) =- i\qb\cdot\kap(t)\cdot\rb_s - 
i\qb\cdot\kap_{tt_0}\cdot(2\bp_s+{\bf F}_s)+ \Omega^{\dagger}(t) - 
\qb\cdot\kap_{tt_0}\cdot\kap^T_{tt_0}\cdot\qb$.
\par
To close Equations (\ref{eom}--\ref{exactmemory}) for the transient density correlator we now make 
an MCT-based approximation to the average in Eq.~(\ref{exactmemory}). Taking care to preserve 
translational invariance of $m_{\qb}(t,t',t_0)$, we obtain 
\begin{eqnarray}
m_{\bf q}(t,t'\!,t_0) \!\!&=& \!\!
\frac{\rho}{16\pi^3} \!\!\int \!\! d\kb 
\frac{S_{k}S_{p} V^{(1)}_{\qb\kb\pb}\,V^{(2)}_{\qb\kb\pb}\Phi_{\kb}(t,t')\Phi_{\pb}(t,t')}
{S_{q}\Gamma_{\bf q}(t'\!,t_0) \Gamma_{\bf q}(t,t_0)}
\notag\\
V^{(1)}_{\qb\kb\pb} &=& \tilde\qb\cdot\big[\,
\hat\kb c_{\hat k} + \hat\pb c_{\hat p} + S_q(\hat\qb c_{\hat q} - \qb c_{q}) 
\,\big]
\notag\\
V^{(2)}_{\qb\kb\pb} &=& 
\qb\cdot\big[ \, 
\kb c_{k} + \pb c_{p} + \kap_{tt_0}\cdot\qb S_q \big\{ c_{k} + c_{p} \notag\\
 &&\hspace*{10mm}- \frac{\rho}{2}\big(c_{q}c_{k}+c_{q}c_{p}+c_{k}c_{p}\big)\big\}
\,\big], 
\label{approxmemory}
\end{eqnarray}
where $\pb=\qb-\kb$;
$\rho=N/V$; $\rho c_q=1-1/S_q$. 
The wavevectors 
$\tilde\qb=\qb - \qb\cdot\kap_{tt_0}$ and
$\hat\qb=\qb - \qb\cdot\kap_{tt'}$ contain shear strains accumulated over different 
temporal ranges.

Equations (\ref{MCTstress},\ref{eom},\ref{exactdecay}) and (\ref{approxmemory}) 
form a closed set of equations to predict the shear stress for arbitrary 
time-dependent shear flows of the form $\kap = \dot\gamma(t)\delta_{ix}\delta_{jy}$. 
Other than $\dot\gamma(t)$,
the only required inputs are density $\rho$ and the static structure factor $S_k$ in the unsheared 
state.
The parametric nature of $t_0$
is made explicit in the MCT approximation (\ref{approxmemory}) where all three 
times enter the vertex functions 
$V^{(1,2)}$
but only 
$t'$ and $t$ enter the correlators. 
For $\dot\gamma=0$ our equations reduce to those of quiescent MCT \protect\cite{MCTequations} 
and for steady shear to those of  
\protect\cite{fuchs_cates_PRL}. 
\par

In developing nonlinear constitutive equations it is helpful to study 
nonlinear step-strain as a benchmark. 
In an idealized step-strain experiment the shear rate is 
given by $\dot\gamma(t)=\gamma\delta(t-t_0)$, which provides a demanding test 
of any constitutive equation. 
For step-strain our approximate  Green-Kubo relation (\ref{MCTstress}) 
reduces to
\begin{eqnarray}
\sigma(t) = \!\!\int\! d{\kb}\bigg\{\!
\frac{\gamma}{16\pi^3}\frac{k_x^2 k_yk_y^{\gamma}}{k k^{\gamma}}
\frac{S'_k S'_{k^{\gamma}}}{S^2_{k^{\gamma}}}\!\bigg\}
\big[ \Phi^{\gamma}_{\kb}(t,t_0)\big]^2\!\!,\;
\label{step_stress}
\end{eqnarray}
where $t>t_0$ and $\kb^{\gamma}=(k_x,k_y - \frac{1}{2}\gamma k_x,k_z)$. 
We have included an additional superscript on the correlator to make explicit 
the strain dependence.
The initial decay rate becomes independent of time, 
$\Gamma_{\qb}^{\gamma} = q^2/S_q + q_x q_y \gamma ( 1+ S_q )/(2S_q)
 + q_x^2\gamma^2/4$,
as the time $t_0$ drops out in favour of an explicit $\gamma$ dependence.
A similar reduction occurs for the memory function, leading us to modify the 
notation, $m_{\bf q}(t,t'\!,t_0)\rightarrow m^{\gamma}_{\bf q}(t,t')$.
The memory function (\ref{approxmemory}) is thus given in step-strain by 
\begin{eqnarray}
m^{\gamma}_{\bf q}(t,t') \!\!&=& \!\!\!
\frac{\rho}{16\pi^3} \!\!\int \!\! d\kb 
\frac{S_{k}S_{p} V^{(1)}_{\qb\kb\pb}\,V^{(2)}_{\qb\kb\pb}
\Phi^{\gamma=0}_{\kb}(t,t')\Phi^{\gamma=0}_{\pb}(t,t')}
{S_{q}\Gamma_{\bf q}^{\gamma}\, \Gamma_{\bf q}^{\gamma}}
\notag\\ 
\label{memory_step}
\end{eqnarray}
with $\tilde\qb \to \qb^\gamma$, hatted variables made unhatted, 
and $\kap_{tt_0} \!\!\to\!\!\kappa^{\gamma}_{ij}\!\!=\!\!\frac{1}{2}\gamma\delta_{ix}\delta_{jy}$ in the 
vertex expressions of Eq.~(\ref{approxmemory}). The strain is zero in the two correlators 
since there is no strain imposed {\em between} $t'$ and $t$; these are quiescent MCT correlators. 
Thus all the $\gamma$ dependence in the memory function stems from 
the presence of $t_0$ in Eq.~(\ref{approxmemory}). 
In Eq.~(\ref{memory_step}) the forces represented by the vertices are strain-dependent 
but relaxation of the structure is $\gamma$-independent \protect\cite{footMRcomment}.
We thus obtain a simplified equation of motion for the correlators 
needed in (\ref{step_stress}). 
At statepoints for which quiescent MCT predicts a glass, 
the memory function remains non-ergodic for all values of $\gamma$.
The equation of motion for $\Phi^{\gamma}_{\bf q}(t,t_0)$ is a linear equation 
in $\Phi^{\gamma}_{\qb}(t,t_0)$ with given non-Markovian memory function. 
Due to the zero duration of the applied strain we can make the replacements 
$\Phi^{\gamma}_{\qb}(t,t_0)\rightarrow \Phi^{\gamma}_{\qb}(t-t_0)$, 
$m^{\gamma}_{\bf q}(t,t')\rightarrow m^{\gamma}_{\bf q}(t-t')$ without 
further approximation.
\par
\begin{figure}
\epsfig{file=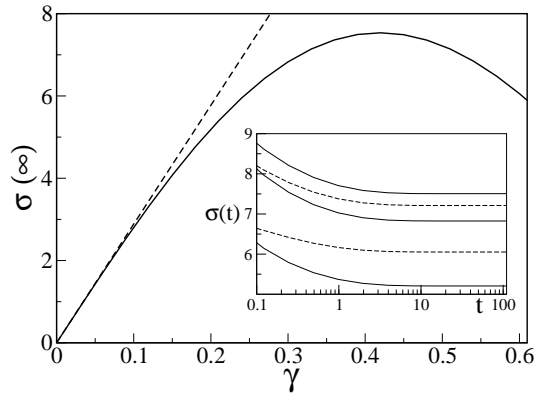,width=5.4cm,angle=-90}
\caption{The long time stress $\sigma(\infty)$ as a function of  strain $\gamma$ 
for a hard sphere glass just above the MCT glass transition ($\phi=0.52$, 
$\varepsilon\equiv(\phi-\phi_c)/\phi_c=0.008$) following a  step-strain.
The dashed line is the linear response result and the solid line is $\sigma(\infty)$
calculated using the isotropic approximation \protect\cite{isotropicapprox}.
The inset shows the decaying stress as a function of time following the step for 
$\gamma=0.2,0.3,0.4$ (full lines, bottom to top) and $\gamma=0.5,0.6$ 
(broken lines,top to bottom). 
}
\label{fig1}
\end{figure}
\par
In practice the presence of anisotropy still poses difficulties for numerical computation; 
specializing to the case of hard spheres
we have therefore solved Eq~(\ref{step_stress}) for the shear stress 
$\sigma(t)$ within the isotropic approximation 
\protect\cite{fuchs_cates_PRL,miyazaki_reichman2,isotropicapprox}.
The $S_k$ input is taken from the Percus-Yevick (PY) theory, 
which yields a glass transition at 
$\phi_{\rm mct}\equiv\pi d^3 \rho/6=0.516$ with our present numerical discretization 
\protect\cite{franosch}. 
In Figure \ref{fig1} we show the long time stress $\sigma(\infty)$ as a 
function of strain amplitude $\gamma$ for a state point 
just above the MCT glass transition ($\phi=0.52$, for fluid states we of course find 
$\sigma(\infty)=0$.)
As $\gamma$ is increased we first obtain the expected linear response behaviour 
before entering a regime of sub-linear increase, indicating the 
onset of plastic flow. 
(We find that a regime of linear response persists even at the critical point 
$\phi=\phi_c$.)
For large $\gamma$ values our calculations yield unphysical negative $\sigma(\infty)$.
This may reflect a shortcoming of MCT, where for large $\gamma$ use of $S_k$ to proxy 
the colloidal interactions becomes questionable.
The inset shows $\sigma(t)$ for various values of $\gamma$. 
In contrast to polymer melts following a step-strain, $\sigma(t)$ for the present hard 
sphere system is not strain factorable \protect\cite{doi_edwards}.

The results (Fig.~\ref{fig1}) are in qualitative agreement 
with recent step-strain experiments on suspensions of PMMA particles above the glass 
transition \protect\cite{pham}.
The experimental data show a peak and a region of negative slope in 
$\sigma(\infty)$ in accord with our findings \protect\cite{footband}. 
The isotropic approximation is known to underestimate the effects of shearing 
\protect\cite{fuchs_cates_PRL}; a fuller treatment might shift the peak 
in $\sigma(\infty)$ from $\gamma\simeq 0.4$ closer to the experimental 
results \protect\cite{pham}, which peak at $\gamma\simeq 0.1$. 
Note also that in the experiments the strain is ramped up 
over some short finite interval, during which additional plastic rearrangment may 
occur. 
Our general expressions (\ref{MCTstress},\ref{eom},\ref{initialdecay},\ref{approxmemory}) 
should capture this although (\ref{step_stress},\ref{memory_step}) clearly do not.

\par
To summarize, for interacting Brownian particles advected by a non-fluctuating shear flow, 
we have generalized the integration through transients formalism of \protect\cite{fuchs_cates_PRL} to 
address arbitrary far-from-equilibrium, time-dependent shearing. 
When complimented with mode-coupling approximations this provides a route to calculating time 
dependent averages in the sheared system. 
As a demonstration of this approach we have presented a Green-Kubo-type relation for the 
shear stress and an 
equation of motion for the transient density correlator, involving a memory function with 
non-trivial time dependence. 
MCT closure of these expressions yielded a first-principles constitutive model for
the shear rheology of dense suspensions close to the glass transition. In step-strain, this
predicts first a linear regime followed by plastic deformation of the glass with a maximum 
in the long-time stress, as seen experimentally.

The formalism excludes extensional flow, and we assumed a spatially homogeneous strain rate 
$\dot\gamma(t)$. 
Nonetheless, the work goes far beyond linear response \protect\cite{gstar}: applying the 
theory to the case of oscillatory shear would predict strain dependent storage and loss 
moduli including `higher harmonic' contributions \protect\cite{expts,wilhelm}. 
Efficient numerical algorithms to tackle the anisotropy and loss of time translation invariance 
are currently under development.
Meanwhile the formal developments presented here form a secure starting point both for more 
complete theories, in which the MCT assumptions might be partially relaxed, and for schematic 
models that simplify the algebra but add extra physics such as anharmonicity, shear-thickening 
and jamming \protect\cite{holmes}. 
\par
We acknowledge the Transregio SFB TR6, EPSRC/GR/S10377 and DFG Vo 1270/1-1 for financial support.
We thank Stefan Egelhaaf, Marco Laurati and Martin Greenall for helpful discussions.


\begin{thebibliography}{}

\bibitem{expts} G. Petekidis, D. Vlassopoulos and P. N. Pusey, 
Faraday Discussions {\bf 123}, 287--302 (2003); 
J. Phys. Cond. Mat. {\bf 16}, S3955 (2004), and references therein.

\bibitem{sgr} P. Sollich {\em et al}, Phys. Rev. Lett. {\bf 78}, 
2020 (1997); 
S. Fielding, P. Sollich and M. E. Cates, J. Rheol. {\bf 44}, 323 (2000); 
M. E. Cates and P. Sollich, J. Rheol. {\bf 48}, 193 (2004).

\bibitem{bouchaud}
J.-P. Bouchaud, J. Physique Paris I, {\bf 2}, 1705 (1992).

\bibitem{cowpaste} S. D. Holdsworth, Trans. Inst. Chem. Eng. A {\bf 71}, 139 (1993).

\bibitem{spingl} L.~Berthier, J.-L.~Barrat and J.~Kurchan, Phys. Rev. E {\bf 61}, 5464 (2000).

\bibitem{fuchs_cates_PRL}
M.~Fuchs and M. E.~Cates, Phys. Rev. Lett. {\bf 89}, 248304 (2002); 
M.~Fuchs and M. E.~Cates, J. Phys. Cond. Mat. {\bf 17}, S1681 (2005);
M. E.~Cates {\em et al}, in {\em Unifying Concepts in Granular Media and Glasses} 
ed. A.~Coniglio {et al} (Amsterdam, Elsevier, 2004) p203.

\bibitem{goetze_sjoegren}
W.~G\"otze and L.~Sjoegren, Rep. Prog. Phys. {\bf 55}, 241 (1992).

\bibitem{pham2002}
K. N.~Pham {\em et al}, Science {\bf 296}, 104 (2002).

\bibitem{vanmegen}
W.~van Megen and S. M.~Underwood, Phys. Rev. Lett. {\bf 70}, 2766 (1993); 
W.~van Megen and S. M.~Underwood, Phys. Rev. E {\bf 49}, 4206 (1994); 
T.~Eckert and E.~Bartsch, Faraday Discussions {\bf 123}, 51 (2003).
 
\bibitem{MCTequations}
W.~G\"otze, {\em Liquids, Freezing and Glass Transition} 
ed. J-P.~Hansen, D.~Levesque and J.~Zinn-Justin 
(Amsterdam: North-Holland, 1991) p287. 


\bibitem{gstar} 
G.~N\"agele and J.~Bergenholtz, J. Chem. Phys. {\bf 108}, 9893 (1998); 
M.~Fuchs and M. R.~Mayr, Phys. Rev. E {\bf 60}, 5742 (1999).

\bibitem{miyazaki_reichman}
K.~Miyazaki and D. R.~Reichman, Phys. Rev. E {\bf 66}, 050501(R) (2002);
K.~Miyazaki, D. R.~Reichman and R.~Yamamoto, Phys. Rev. E {\bf 70}, 
011501 (2004).

\bibitem{miyazaki_reichman2}
K. Miyazaki {\em et al}, cond-mat/0509121.

\bibitem{kss} V.~Kobelev and K. S.~Schweizer, Phys. Rev. E {\bf 71}, 021401 (2005).

\bibitem{holmes} C. B. Holmes {\em et al}, J. Rheol. {\bf 49}, 237 (2005).


\bibitem{dhont}
J. K. G.~Dhont, {\em An Introduction to the Dynamics of Colloids}, 
(Amsterdam, Elsevier, 1996).

\bibitem{doi_edwards}
M.~Doi and S. F.~Edwards, {\em The Theory of Polymer Dynamics}, 
(Oxford University Press, 1989). 

\bibitem{normalstress}
Normal stresses can be found the same way by choosing 
$f = (\sigma_{xx}-\sigma_{yy})/V$, {\em etc.} and deriving analogs to Eq.~(\ref{exactstress}).




\bibitem{ch}
B.~Cichocki and W.~Hess, Physica A {\bf 141}, 475 (1987).



\bibitem{footMRcomment} 
Note that the approach of \protect\cite{miyazaki_reichman,miyazaki_reichman2} neglects the strain 
accumulated between $t_0$ and $t'$ and gives a strictly linear stress-strain relation
when applied to step-strain. 

\bibitem{isotropicapprox}
In the isotropic approximation
we replace both the numerator and denominator in 
(\ref{memory_step}) by the angle average over $\qb$ to obtain 
an isotropic memory function $m_q^{\gamma}(t,t')$. 
If we also isotropize the initial decay rate then we obtain an isotropic 
correlator. This is then used in Eq. \ref{step_stress}.

\bibitem{franosch}
We employ the same numerical discretization as
T.~Franosch {\em et al} 
Phys. Rev. E {\bf 55}, 7153 (1997). 

\bibitem{pham}
K. N.~Pham {\em et al} Europhys. Lett. {\bf 75}, 624 (2006); 
W. C. K.~Poon (private communication). 

\bibitem{footband}
Note that $d\sigma/d\gamma <0$ for an elastic system would cause a mechanical 
(static shear-banding) instability. However, since $\gamma$ denotes an initial strain 
and $\sigma(\infty)$ a final stress, this might not arise under plastic flow conditions. 
No shear banding is reported in \protect\cite{pham}.

\bibitem{wilhelm} M.~Wilhelm, Macromol. Mater. Eng. {\bf 287}, 83 (2002).
\end{thebibliography}
\end{document}